# Comment: Struggles with Survey Weighting and Regression Modeling

**Robert M. Bell and Michael L. Cohen**

Andrew Gelman's article "Struggles with survey weighting and regression modeling" addresses the question of what approach analysts should use to produce estimates (and associated estimates of variability) based on sample survey data. Gelman starts by asserting that survey weighting is a "mess." While we agree that incorporation of the survey design for regression remains challenging, with important open questions, many recent contributions to the literature have greatly clarified the situation. Examples include relatively recent contributions by Pfeffermann and Sverchkov (1999), Graubard and Korn (2002) and Little (2004). Gelman's paper is a very welcome addition to that literature.

There are some understandable reasons for the current lack of resolution. First, U.S. federal statistical agencies have been historically limited by their mission statements to producing statistical summaries, primarily means, percentages, ratios and cross-classified tables of counts. This is one explanation for why Cochran (1977) and Kish (1965) devote the great majority of their classical texts to these estimates. As a result, the job of using regression and other more complex models to learn about any causal structure underlying these summary statistics was generally left to sister policy agencies and outside users.

However, things are changing. The federal statistical system (whether it likes it or not) is becoming more involved with complex modeling. This includes small-area estimation (e.g., unemployment estimates and census net undercoverage estimates) and research into models combining information from surveys with administrative data. (There will also likely be increased demands to use data mining procedures on federal statistical data.) This relatively new development has likely motivated several of the recent contributions on how to account for the sample design in complex models. Therefore, Gelman's article and the resulting discussion come at an important time.

Another reason for the failure to resolve this class of problems is that this general issue is not easy. Attempts to resolve this problem raise a number of clashing perspectives, including: (1) whether to be model-based or design-based in one's inference, (2) whether to take a Bayesian or a frequentist view, (3) whether one's inference should be conditional on (some of) the observed values of the design variables and other auxiliary data that one might have for the full population, (4) whether one evaluates a procedure based on its small-sample performance or its asymptotic properties, and (5) whether one wants an algorithm specific to a particular regression model or something more omnibus.

A variety of general schemes have been proposed to deal with this hard problem, and several of them can be expressed as members or mixtures of the following pure strategies: (1) use an unweighted analysis of the collected data, which is a pure model-based perspective assuming the model is correct for the entire (super) population, (2) use the inverses of the sample selection probabilities as weights, which derives from a pure design-based perspective and is therefore not dependent on model-based assumptions either, and (3) include the survey design in the model as predictors (Little, 2004). The last strategy, for instance, would make sense if it was obvious that separate models were needed for subgroups defined by the survey variables. Gelman's paper represents a mixture of strategies (2) and (3).

It is useful to take a closer look at the second example in Section 1.4 of Gelman's article, which addresses the bias of the race coefficient for predicting log income when the sample is unrepresentative of the population in terms of gender. Like Gelman, we are viewing the problem as one of estimating the "so-called" census regression coefficient, which in this case is the mean log income for whites minus the mean log income for nonwhites in the finite population. Some algebra shows that conditional on the population margins and assuming that data are missing at random, the bias of the race coefficient in a simple unweighted regression of log income on race is approximately proportional to the product of two factors: the proportion of males in the sample minus the proportion in the population and the race–gender interaction for the population. In this simple setting, the bias is equivalently correctable either by weighting the simple regression or with the model-based algorithm outlined by Gelman. Given the large interaction stipulated in the article, it is imperative that the bias be corrected, assuming a nontrivial deviation in terms of gender between the sample and the population.

*Robert M. Bell is Member, Statistics Research Department, AT&T Labs–Research, 180 Park Avenue, Florham Park, New Jersey 07932, USA e-mail: rbell@research.att.com. Michael L. Cohen is Study Director, Committee on National Statistics, National Academies, Room 1135 Keck Center, 500 5th St., N. W., Washington, District of Columbia 20001, USA e-mail: mcohen@nas.edu.*

However, that may not be true in general. Whether one should try to correct for bias should also take into account the impact on the variance. Either weighting or modeling inflates the variance of the resulting estimate for the race coefficient. Unlike the bias, the added variance depends only on how much the distribution of gender in the sample differs from that in the population and not on the size of the interaction. When the true interaction is very small, the mean-squared error will increase if we try to correct for the bias either through weighting or modeling. On the other hand, for sufficiently large interaction effects, the correction decreases mean-squared error. The sample imbalance does not affect whether one is better off correcting, but only the magnitude of the expected benefit or harm from the correction.

In general, the size of the true interaction that implies one should correct for bias is on the order of the empirical uncertainty associated with the estimated interaction, so it is impossible to conclude with much confidence that correcting for bias is the wrong strategy. Consequently, it is a no brainer to simply correct for the bias by either weighting or modeling, unless one has strong prior evidence that the interaction truly is very small.

However, surveys often have many potential stratifying variables, perhaps including some like state, with dozens of levels. For example, consider a longitudinal study where we would like the follow-up sample after nonresponse to represent the baseline sample. There may be dozens or even hundreds of variables on which we would like to balance. Even with a few variables, it quickly becomes impractical either to form a complete cross-classification for weighting or to fit a model that represents all interactions of the original model covariates with variables related to the sample design. Some sort of compromise is imperative, and the question is how to choose it.

Survey practitioners use all sorts of compromises: at the crudest level, cross-classification while omitting some variables and/or collapsing values for other variables; raking or propensity scores weights based on logistic regression of response at follow-up using selected interactions; and tools like weighting cells and weight trimming to control the variability of estimates. Modelers have an equally varied assortment of options at their disposal.

Does it matter whether one uses weights or a model-based approach? As Gelman shows, there is a correspondence between the corrections available by modeling versus weighting, so either path can work well. What matters most, we believe, is that decisions about which variables and interactions to use should be informed by the interactions that actually predict the outcome. In particular, even though weights can be created without even looking at the outcome, the best weights are likely to be ones that were informed by an appropriate model.

Gelman's hierarchical regression model approach has some very appealing features. It supports the use of rich models of the dependent variables while at the same time reducing the chance of overfitting. Rather than treating interaction terms as either in or out, shrinking estimated interactions adaptively often improves predictive accuracy, and, most likely, bias correction. These models also provide a principled basis for inference, which is hard to argue if "design-based" weights are chosen based on a modeling exercise. Finally, the paper helps to clarify the relationship between modeling and weighting for bias correction, by demonstrating that the modeling methodology implies the use of weights. This is important because weighting offers several practical benefits. These include (a) the ability to use standard software routines, (b) avoidance of the need to fit large models with many interactions (fixed and/or random effects) every time one wants to estimate even the simplest new regression model, and (c) the potential to provide for users of data from a government agency a simple way to produce near-optimal results.

Point (c) is somewhat Pollyannish and in need of some amplification. The ideal weights would vary from regression to regression, and the use of these weights would create a lot of work and would greatly complicate comparisons across analyses. To the extent that constant weights were proposed for use, one would want the weights to be such that they would work reasonably well across a range of potential regression analyses. Which terms to include in either a design-based or a model-based solution should depend on the size of various interactions on the dependent variables of interest. Unfortunately, a good set of weights for one regression analysis may be quite poor for another one. However, possibly the outcome variables could be grouped and a set of weights identified that work reasonably well for the entire group of variables.

We hope that researchers continue to investigate, as Gelman has suggested, the relationship between weighting and modeling to try to develop approaches that enjoy the best of both worlds, in particular that are omnibus for a variety of estimands of interest. Returning to the federal statistical system, given its need to produce a large number of estimates, often disaggregated demographically and geographically, for its large and diverse user community, there is an important advantage to more general-purpose and easy-to-apply methods.

Finally, we have a couple of questions or issues that could use further work or explication:

- Gelman's method for estimating and producing inferences for census regression parameters relies on a hierarchical regression model, so it is important to understand the quality of fit of that model. However, for hierarchical regression models estimated using data from a complex sample, notions of standardized residuals and leverage and their use in assessing linearity, influence, variance heterogeneity, and so on, are quite complicated. Further, even with adequate diagnostics for hierarchical regression models, those diagnostics will not assess the influence of particular data points on the census population regression estimates. It would be valuable to investigate these issues further. (Initial efforts toward incorporating sample weights in diagnostic plots have been taken by Korn and Graubard, 1995.)
- Finally, although Gelman focuses on the goal of estimating linear regression parameters, he mentions that his techniques may extend to logistic regression. Modern data analysis makes use of a much wider variety of techniques, as found, for example, in Hastie, Tibshirani and Friedman (2001). For example, in classification and regression trees, the parameters play a very different role, and iterative steps are used to "grow" the tree. It is unclear how either a model-based or a weighting approach should be used in either growing classification or regression trees, or in assessing their performance on a training sample that was collected from a complex sample design. Research on the interface of these problems would be valuable.

In summary, Gelman's research makes very valuable contributions to the question of how to carry out regression modeling from complex samples. Clearly, as Gelman has stated, more work is needed in this area.

## ACKNOWLEDGMENT

We greatly appreciate Phil Kott's critique of an earlier version of this comment.